**Highlights**

- Particle-turbulence interaction in high-*Re* turbulent boundary layer is studied via a large field-of-view 2D PIV/PTV.
- A critical layer is found to partition the kinematical characteristics of the motion of solid particles with high inertia.
- The critical layer might be a good indicator of the upper bound of the particle saltation process.
- The small-scale turbulent motion, i.e. the near-wall burst events, of the gas phase is attenuated by solid particles.
- Large-scale motions (LSMs) in the log layer of the turbulent boundary layer are found to play important role in the particle saltation process.

# Particle-turbulence interaction in High-Reynolds-number Sand-laden Turbulent Boundary Layer

Hang-Yu Zhu$^a$, Chong Pan$^{a,*}$, Jin-Jun Wang$^a$, Yi-Rui Liang$^b$

$^a$*Key Laboratory of Fluid Mechanics (Beijing University of Aeronautics and Astronautics),*
*Ministry of Education, Beijing 100191, China*
$^b$*Key Laboratory of Mechanics on Disaster and Environment in Western China (Lanzhou University),*
*Ministry of Education, Lanzhou 730000, China*

**Abstract:** Simultaneous two-phase particle image/tracking velocimetry (PIV/PTV) measurement is conducted on particle-laden turbulent boundary layer (TBL) over a horizontal smooth-flat-plate. The relatively high Reynolds number ($Re_\tau$=5500 based on friction velocity $u_\tau$ and boundary layer thickness $\delta$) wind-sand TBL with large field-of-view (FOV) is experimentally explored. With four high-resolution CCD cameras arranged along the flow direction, this experiment can resolve a wide range of the spectrum ranging from small-scale energetic eddies to large-scale motions (LSMs) in TBL. Dilute desert sand grains with median diameter of 203$\mu m$ and bulk volume fraction of $O(10^{-5})$ are used as discrete phase. Improved phase separation and discrete particle matching method are developed for two-phase velocity measurement. The results presented here provide new information concerning the effect of high-inertia particles with dilute concentration on wall-bounded turbulence, especially at high $Re$. The presence of sand grains attenuate turbulence fluctuation by suppressing small-scale sweep-ejection cycle in the near-wall region, and this suppression effect intensifies as the particle concentration increases. A critical layer ($y/\delta$=0.12 or $y^+$=670) is found to partition the streamwise evolution of both the local concentration and the streamwise mean velocity of the sand grains into a spatial developing near-wall region and a quasi-parallel outer region. In addition, at this layer a balance of the strength and probability between the sweep and ejection events of sand grains is reached. Such a critical layer might be a good indicator of the upper bound of the particle saltation process, in which LSMs are believed to play a significant role.

**Key words:** Particle-turbulence interaction; Sand-laden turbulent boundary layer; Large-scale motions; Particle image/tracking velocimetry

## 1. Introduction

Turbulent particle-laden flows are widely encountered in diverse environmental and industrial processes, such as sand storm, coal combustion, fluidized beds and pneumatic conveyance. These two-phase flows can be roughly divided into three regimes: one-way, two-way and four-way coupling, each of which has different degree of particle-turbulence interaction (Elghobashi 1994). In the first regime, the effect of particles on the fluid flow is too small to be considered; while the particle-turbulence interaction in the rest two regimes are non-negligible. Furthermore, the effect of inter-particle collisions becomes prominent in the four-way coupling regime.

---

* Corresponding author: Dr. Chong Pan
Tel: +86-10-82338069
Email: panchong@buaa.edu.cn

In the fundamental research aspect, two key issues, i.e. turbulence modification with the presence of particles and particle transport by turbulent flows, are directly related with the dynamics and characteristics of turbulent particle-laden flows. Existing experiments and numerical simulations have greatly contributed to the understandings on these two issues. However, since a number of non-dimensional parameters, such as particle Reynolds number $Re_p$, particle Stokes number $St_p$, particle volume fraction $\Phi_V$, particle-to-fluid density ratio $\rho_r$, particle-to-turbulence length-scale ratio, etc., are involved (Saber et al. 2016), a full description of the underlying physics and mechanisms of this complicated problem is still lacking.

Both turbulence augmentation (Sato and Hishida 1996; Suzuki et al. 2000; Kiger and Pan 2002; Goswami and Kumaran 2011) and turbulence attenuation (Rogers and Eaton 1991; Li et al. 2012; Li et al. 2016, 2018) have been identified in the previous studies, and various criteria were proposed to account for such difference in turbulence modification. Among them, the ratio of the particle size to the characteristic turbulent length scale was the first one being recognized (Tsuji et al. 1984; Gore and Crowe 1989; Pan and Banerjee 1996). Meanwhile, Hetsroni (1988) suggested that $Re_p$ was another critical parameter in a sense that particles with large $Re_p$ enhanced the turbulent fluctuation by shedding wakes into the flow. Kulick et al. (1994) and Zhao et al. (2013) further showed that the degree of turbulence attenuation increased monotonically with $St_p$ (or particle response time). Tanaka and Eaton (2008) derived a new dimensionless particle moment number $Pa$ from the particle-laden Navier-Stokes equations to categorize the turbulence modification.

Despite of the knowledge on the role of these critical parameters, the boundary between turbulence augmentation and attenuation is still confusing, and some conflicting results have been published. In the numerical study of a particle-laden vertical channel flow, both Li et al. (2001) and (Dritselis and Vlachos (2008), 2011)) reported that the addition of inertial particles suppressed the streamwise turbulent fluctuation in the near-wall region but enhanced that in the outer region. Nevertheless, this observation was opposite to the studies of (Li et al. (2012); Li et al. (2016)) via either experiment or simulation. Furthermore, Li et al. (2001) and Dritselis and Vlachos (2011) found that particles with higher inertia had smaller impact on turbulent velocity fluctuations, but both Kulick et al. (1994) and Zhao et al. (2013) showed that the turbulence suppression intensified with the increase of $St_p$.

Note that even in the one-way coupling regime, remarkable turbulence modification can be still observed, which might be attributed to the distortion of turbulent structures induced by particle preferential concentration (Tanière et al. 1997; Wu et al. 2006; Lian et al. 2013). On the other hand, Goswami and Kumaran (2011) found that the polydispersity of the particle size led to the variation of the particle terminal velocities, which in turn enhanced the inter-particle collisions and eventually changed the turbulent fluctuation level. Nasr et al. (2009) and Li et al. (2018) reported a similar effect of inter-particle collisions on turbulence modification. Therefore, it is reasonable to infer that the turbulence modification is a result of not one single critical parameter but the combined effects of multiple competing factors. This forms the direct impetus for a continuing investigation on this issue (Balachandar and Eaton 2010).

As for the aspect of particle transport, it is widely accepted that turbulent coherent structures affect particles' entrainment and deposition behavior to a certain degree. In the near-wall region of wall-bounded turbulence, the bursting events were found to be responsible for the wall-normal motion of inertial particles (Kaftori et al. 1995; Marchioli and Soldati 2002; Hout 2011, 2013), and particles tended to accumulate into streamwise low-speed streaks (Fessler et al. 1994; Kaftori et al. 1995; Marchioli et al. 2003; Håkansson et al. 2013). An interesting phenomena that raised more attention recently is that the properties of turbulent structures, i.e. strength and length/time scales, are also susceptible to particle motions. Such particle-induced structural modifications have been observed in busting events (Rashidi et al. 1990; Pan and Banerjee 1996; Kaftori et al. 1998; Li et al. 2012), inner-layer quasi-streamwise vortices (Dritselis and Vlachos 2008) and low-speed streaks (Li et al. 2016). However, similar to the turbulence modification, remarkable conflicts in the previous studies on this issue still exist. For instance, Dritselis and Vlachos (2008) found that heavy particles increased the streamwise extent of quasi-streamwise vortices, while Li et al. (2012) and Li et al. (2016) reported that heavy particles preferred to reduce the coherency of near-wall vortical structures.

Besides the above conflicting observations, there are several challenges that keep us from establishing a sophisticated description on the particle-turbulence interaction. First of all, most of the reported studies focused on particle-laden turbulence either in homogeneous and isotropic turbulent flows or in parallel flows (e.g. pipe or channel flows), while studies on spatial developing turbulent boundary layers (TBL), which are more close to practical applications, were comparably rare. Among these few studies, early experiments (Rogers and Eaton (1991); (Kaftori et al. (1995), 1995)) and Tanière et al. (1997)) and recent direct numerical simulations (DNS) (Li et al. (2016), 2018)) were mainly limited to low Reynolds number ($Re$) condition. Secondly, turbulence modulation by high-inertia particles were less addressed in the past, so did the effect of gravity (Marchioli et al. 2007; Nilsen et al. 2013; Mathai et al. 2016) and particle-particle or particle-wall collisions (Li et al. 2018). Lastly, in high-$Re$ scenario, large-scale motions (LSMs) and very large-scale motions (VLSMs) significantly contribute to the production and transport of turbulent kinematic energy (TKE) and Reynolds shear stress (RSS) (Monty et al. 2007; Mathis et al. 2009; Wang and Zheng 2016; Deng et al. 2018). How do these large-scale structures affect the particle dynamics? Do they play a determining role in the vertical transport of particles in the outer layer? Answers to these questions might contribute to a new perspective on large-scale coherent motion of particles, which is embodied as sand-grain transport over long distance in a sand storm (Wang and Zheng 2016; Wang et al. 2017; Zhang et al. 2018).

In order to cope with these challenges, the present work carried out an experiment in a dilute particle-laden high-$Re$ TBL via simultaneous particle image velocimetry (PIV) and particle tracking velocimetry (PTV) measurement. The primary aim is to study the interaction between heavy particles, whose gravity effect cannot be neglected, and canonical high-$Re$ TBL, in which LSMs and VLSMs might play a crucial role. In the following, §2 describes the experimental set-up and the PIV/PTV measurement details. §3 discusses the streamwise evolution of particle properties to form a partition of particle behaviors at different flow layers. The kinematic statistics of both the gas phase and the particle phase is presented in §4, with the emphasis on the turbulence modification under the effect of particles. Both quadrant analysis and conditional average analysis are taken in §5 to delineate the correlation between the particle motions and large-scale structures/events. Additional discussion and concluding remarks are finally drawn in §6.

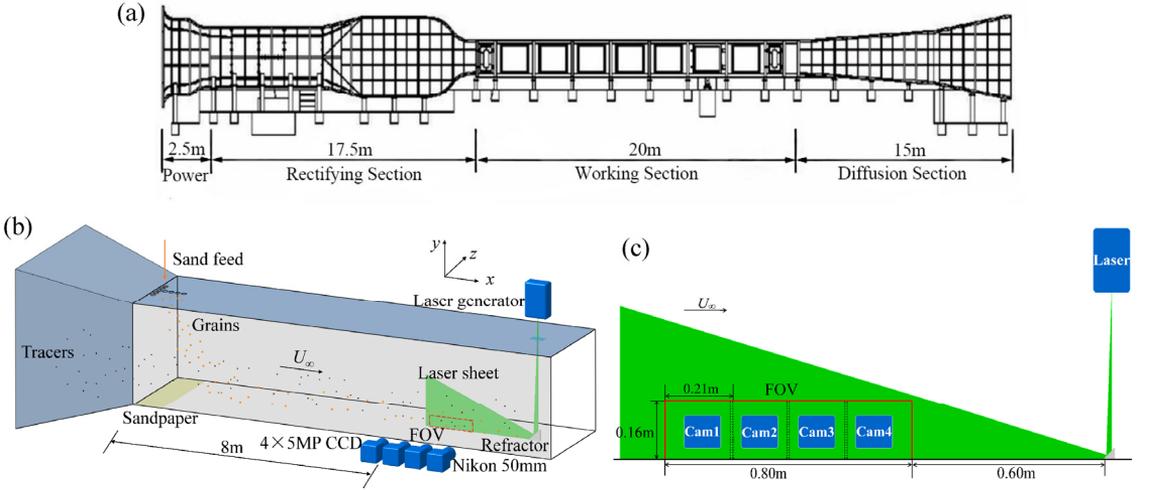

Figure 1. Illustration of the experimental setup. (a) sketch of the wind tunnel configuration; (b) sketch of both the particle-feeding system and the PIV system; (c) enlarged view of the arrangement of the laser, CCD camera and the field of view.

## 2. Experimental setup and methodology

### 2.1 *Experiment Facility*

The present experiment was conducted in a blow-type multi-functional environmental wind tunnel in Lanzhou University, China. As shown in figure 1(a), the axial fan of this tunnel locates next to the tunnel inlet, after which the settling camber with contraction ratio being 6.44:1 contains one set of honeycomb and three sets of stainless mesh screen. The following test section has a rectangular cross-section of height 1.45m and width 1.3m, and is 20m in length. The top wall has a minor expansion angle of 0.8° to minimize the effect of the growing boundary layer thickness on the local pressure gradient. The variation of the free-stream velocity along the axial line of the test section is less than 1% over 4 m. The floor and sidewalls of the test section are made of Plexiglass for full optical access, and can be regarded as hydraulically smooth.

The free-stream wind velocity $U_\infty$ of this facility varies from 4 m/s to 30m/s. In the present experiment, $U_\infty$ was set to be 7.5m/s, and the corresponding turbulence intensity $Tu$ was about 1%. As illustrated in figure 1b, a strip of sandpaper with grit size of 60 and length of 0.8 m was pasted on the floor shortly after the test section inlet. With $U_\infty$=7.5m/s, the tripped boundary layer above the floor becomes fully developed turbulence after 3 m in the downstream. In the following, capitals are used for time-averaged mean values, lowercase symbols for instantaneous values, and subscript of *rms* for the root-mean-squared fluctuation intensities. In addition, variables with superscript of + indicate inner-scale normalization using the friction velocity $u_\tau$ and the kinematic viscosity $v$ of the air. The streamwise, wall-normal and spanwise coordinates, as shown in figure 1b, are indicated as $x$, $y$, and $z$, respectively, and the corresponding velocity components are $U$, $V$, and $W$. The subscripts $f$ and $p$ denote variables of the continuum fluid phase and the discrete particle phase, respectively.

Figure 2 shows the inner-scaled wall-normal profile $U^+(y^+)$ of the mean streamwise velocity of the baseline single-phase TBL (denoted as Case S in the following) at 8 m downstream the transition strip. A standard 2D PIV system with spatial resolution of 28 wall units/vector was used to obtain such a profile. The $U^+(y^+)$ profile measured by the PIV system with an array of 4 CCD cameras, which will be described later, is also supplemented for a cross validation. The characteristic parameters of this baseline TBL are summarized in Table 1. The friction Reynolds number based on the boundary layer thickness $\delta$ and the friction velocity $u_\tau$ was $Re_\tau = u_\tau \delta/\nu = 5500$, with $\delta = 0.30$ m being determined by the modified Clauser fit with fit range from $y^+=60$ to $y^+=850$, $u_\tau = 0.266$ m/s assessed by the Clauser chart with log-law constants of $\kappa=0.384$ and $B=4.12$ (Marusic et al. 2013). The boundary layer shape factor was $H=1.36$, close to the prediction of the empirical correlation suggested by Marusic et al. (2010).

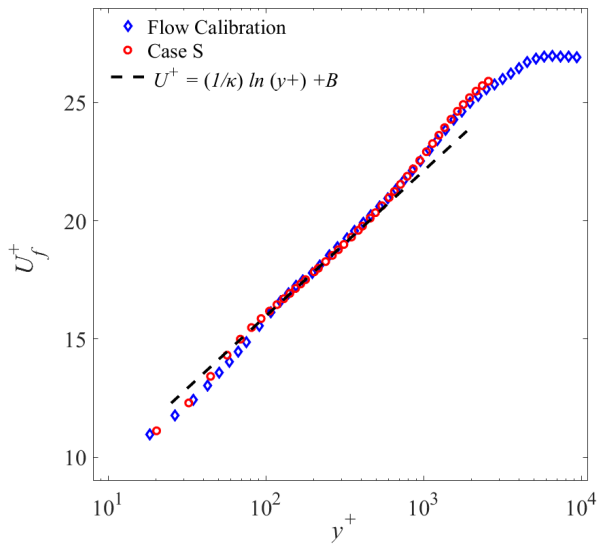

Figure 2. Wall-normal variation of the mean streamwise velocity of the TBL of the baseline Case S. Note that the spatial resolution of the present PIV measurement was 28 wall units/vector, so that the statistics below $y^+<20$ (or $y<1$mm) are unreliable and are thus not shown.

| $\rho_f$ (kg/m³) | $\delta$ (m) | $U_\infty$ (m/s) | $u_\tau$ (m/s) | $H$ | $Re_\tau$ | $Re_\theta$ | $\tau_l$ (s) | $\tau_L$ (s) |
|---|---|---|---|---|---|---|---|---|
| 1.21 | 0.30 | 7.5 | 0.266±0.002 | 1.36 | 5500 | 15700 | 3.24×10⁻³ | 0.4 |

Table 1. Characteristic parameters of the single-phase turbulent boundary layer of Case S

| $\rho_p$ (kg/m³) | $\rho_r$ | $d_p$ (μm) | $d_p^+$ | $\tau_p$ (s) | $Re_p$ | $V_s$ (m/s) | $St_l$ | $St_L$ |
|---|---|---|---|---|---|---|---|---|
| 2600 | 2148 | 203 | 3.6 | 0.210 | 7.5 | 2.0 | 64.8 | 0.54 |

Table 2. Properties of tested sand grains

## 2.2 Characteristics of sand grains

In the particle-laden measurement, sieved desert sand grains were released from the ceiling at the test section inlet through a feeding system whose sand-grain flux can be precisely controlled. As summarized in Table 2, the mean density of sand grains was about $\rho_p=2600$ kg/m³, and the density ratio of the sand grains to the air was $\rho_r=\rho_p/\rho_f=2148$. The sand-grain diameter, measured by a laser scattering particle analyzer (Microtrac S3500), presents a distribution that is slightly deviated from Gaussian distribution (as shown in figure 3).

The mean diameter was $d_p$=203 $\mu m$ (or $d_p^+$=3.6) and the standard deviation of the diameter is 89 $\mu m$.

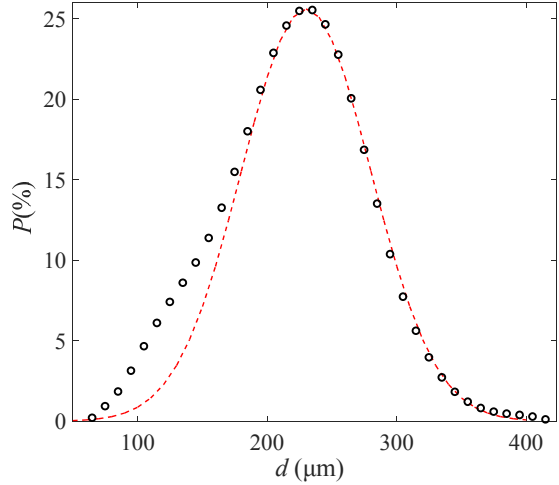

Figure 3 Size distribution of tested sand grains. The red dashed line denotes a Gaussian distribution.

The particle Reynolds number, based on the maximum slip velocity $u_r=|u_p-u_f|$ (Kiger and Pan 2002), was $Re_p= d_p u_r/\nu$=7.5, larger than unity. This means that the flow around sand grains lies in the non-Stokes regime, so that the effect of particle inertia cannot be neglected. According to Coulson et al. (1999) and Kumaran (2003), the corrected particle relaxation time $\tau_p$ incorporating the particle inertia effect can be formulated as

$$\tau_p = \frac{\rho_p d_p^2}{18\nu\rho_f\left(1+0.15\,\mathrm{Re}_p^{0.687}\right)}. \tag{0.1}$$

The particle terminal settling velocity $V_s$ in a quiescent fluid can then be calculated as

$$V_s = \tau_p g, \tag{0.2}$$

with $g$ the gravitational acceleration. The particle Stokes number $St_p$ measures the ratio of $\tau_p$ to one characteristic time scale $\tau_f$ of the single-phase turbulence, i.e.

$$St_p = \tau_p/\tau_f. \tag{0.3}$$

According to Tanière et al. (1997), $\tau_f$ takes either the Kolmogorov time scale $\tau_l$ or the time scale of the largest turbulent eddies $\tau_L$, both of which can be estimated as

$$\tau_l \approx \left(\nu\delta/(0.1U_\infty)^3\right)^{1/2}, \tag{0.4}$$

$$\tau_L \approx \delta/(0.1U_\infty). \tag{0.5}$$

The magnitudes of $\tau_l$ and $\tau_L$ in the baseline Case S are listed in Table 1. The corresponding particle Stokes number were $St_l$=64.8 and $St_L$=0.54. This implies that the wind-blown sand-grain motions are neither in a pure suspension state (i.e. $St_l$<<1) nor a pure saltation state (i.e. $St_L$>>1). Instead, it is subject to the effect of both the fluid turbulence and the particle inertia.

### 2.3 Two-phase flow field measurement

A 2D PIV system was used to measure the two-phase flow field in the symmetric streamwise—wall-normal plane of the TBL at 8 m downstream the tripping strip. In this PIV system, a dual-head Nd: YAG laser with wavelength of 532 nm and energy output of 200 mJ/pulse was used as the light source. The laser beam was shaped into a planar sheet via a set of optical lens to illuminate the symmetric plane of the test section with a thickness of 1 mm. In order to minimize the wall reflection that affects the near-wall measurement, the laser sheet was directed along the upstream direction via a refractor mirror being placed at 0.6 m downstream the measurement region. DEHS droplets with diameter of 0.5~5μm, generated by a pressurized liquid-droplet seeding generator, were released into the free stream before the honeycomb to serve as tracer particles of the gas phase.

| | $U_\infty$ (m s$^{-1}$) | $u_\tau$ (m s$^{-1}$) | $F_s$ (Hz) | $\Phi_1$ (×10$^{-5}$) | $\Phi_2$ | $N_s$ | $N_p$ |
|---|---|---|---|---|---|---|---|
| Case S | 7.5 | 0.266 | 5 | 0 | 0 | 5000 | 0 |
| Case T1 | 7.5 | 0.264 | 5 | 7 | 0.15 | 9000 | 25, 797, 863 |
| Case T2 | 7.5 | 0.264 | 5 | 11 | 0.23 | 9000 | 53, 024, 603 |

Table 3. Characteristic parameters of three tested conditions

As illustrated in figure 1(b), four synchronized double-exposure CCD cameras (IMPERX ICL-B2520M) with resolution of 2058 × 2456 pixel and bit length of 12 bit were aligned along the streamwise direction. Through four Nikon 50mm $f$/1.8D lens, they jointly imaged the laser-sheet-illuminated domain with a field of view (FOV) of about 0.80 ($x$) × 0.16 ($y$) m$^2$. The individual FOV of neighboring CCD cameras had an overlap of 6 mm in $x$ direction to facilitate the velocity-field stich in the post-process procedure. Note that the whole FOV domain corresponds to 2.67$\delta$×0.53$\delta$, large enough to accommodate the streamwise extend of log-layer LSMs without relying on a Taylor frozen spatial-temporal transformation. Meanwhile, the optical magnification is about 0.084 mm/pix, small enough to well resolve the energetic small-scale eddies. Special arrangement of the set of optical lens was made to guarantee that the laser sheet thickness was fixed as about 1 mm in the whole FOV domain.

In addition to the baseline single-phase Case S, particle-laden flows with two different sand-gain flux were tested (denoted as Case T1 and Case T2 hereinafter). As summarized in Table 3, their particle bulk volume fractions, measured by an optical method to be discussed in §2.4, were $\Phi_1$=7 ×10$^{-5}$ and 11×10$^{-5}$, and the corresponding mass loading ratios were $\Phi_2$=$\rho_r\Phi_1$ = 0.15 and 0.23. Due to the dilute particle density and the relatively large wind speed, no severe particle deposition on the wall is seen, which can be illustrated in figure 4. This means that the smooth-wall condition of the TBL is roughly hold for three tested cases.

The sampling repetition rate of all three tested cases was 5 Hz, and the time interval between two straddle frames was fixed to be 100 $\mu s$, yielding a maximum offset of the tracer particle being about 9 pixels (or 0.7 mm). The sampling ensemble in Case S contains $N_s$=5000 image pairs, corresponding to a sampling duration of about $\Delta T$=1000s or $\Delta T^+ \approx$ 4.7×10$^6$, while it enlarges to $N_s$=9000, $\Delta T$=1800s and $\Delta T^+ \approx$ 8.5×10$^6$ in Case T1 and T2 to guarantee a convergence of the velocity statistics of the particle phase. As shown in Table 3, due to the dilute particle concentration, the magnitude of $u_\tau$ only slightly changes in two particle-laden cases, so does $\delta$. Therefore, in the following analysis both the inner- and the outer-scaling of two phases' statistics in all three cases will be non-dimensionalized by the length scales of Case S.

For the single-phase flow (Case S), the state-of-the-art cross-correlation algorithm with multi-resolution iteration and window deformation (Scarano and Riethmuller 2000) was used for 2D velocity field calculation. The interrogation window of the final pass had a size of 32×32 pixels$^2$ and an overlap ratio of 75%, resulting in a spatial resolution of about 1.33 mm/vector or 33 wall units/vector. The relative error of the gas-phase velocity measurement was estimated to be around 1%.

For the particle-laden flows (Case T1 and T2), PIV and PTV were combined to extract the gas- and particle-phase velocity from image pairs containing both tracer particles and sand grains. The image processing procedure contains two steps: phase separation and particle matching. Due to the remarkable difference between the size and brightness of two kinds of particles, which can be clearly seen in figure 4(a), the conventional median filtering method (Kiger and Pan 2000; Hwang and Eaton 2006) was applied for phase separation. In short, one particle image (figure 4a), whose background was subtracted in advance, was convoluted with a median filter with big kernel size to block the tracer particles (see figure 4b). The original image was again filtered by another median filtering with smaller kernel size. The yielded image component was then subtracted from the original one to remove big sand grains (see figure 4c). According to Kiger and Pan (2000), the size of the filter is a critical parameter for the success of the phase separation. After a trial-and-error test, the size of the smaller filter was chosen to be 3×3 pixel, equivalent to the typical size of the tracer particles in the particle image, and that of the bigger one was 7×7 pixel, slightly smaller than the averaged sand-grain size (around 9×9 pixel). The effectiveness of the selected median filter can be demonstrated by figure 4.

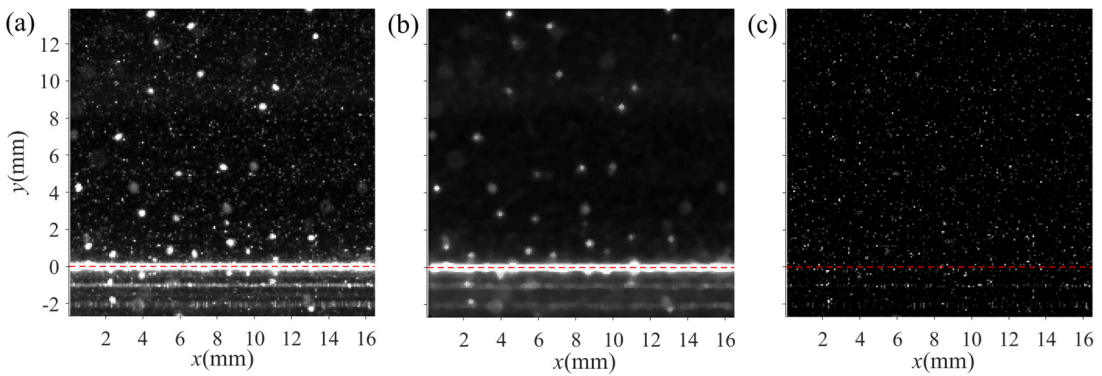

Figure 4. Illustration of the phase separation of two-phase particle images. (a) Image with two-phase particles; (b) image with sand grains; (c) image with tracer particles. The red dashed lines indicate the tunnel floor.

After phase separation, image pairs with tracer particles were processed by the conventional PIV algorithm described in the above. While those with sand grains were analyzed by an in-house hybrid PTV-PIV algorithm. In this algorithm, PTV matching using the minimum offset criteria (Baek and Lee 1996; Ohmi and Li 2002) only provides a rough estimation of the displacement between one pair of object and candidate sand grains in two straddle frames, based on which PIV matching refines the displacement to a sub-pixel level by cross correlating the template window centering around the object sand grain in the first frame and the target window containing the candidate in the second frame. This hybrid algorithm minimizes the contamination from the inaccurate particle-center identification usually encountered in classical PTV algorithm. Additionally, a cross validation is taken by performing both forward matching (from the first frame to the second one) and backward matching (from the second frame to the first one) to reduce outliers caused by particle missing. In the present case, the template window had size of 14×14 pixel, and the target window was 2.5 times larger. The uncertainty of the sand-grain matching was estimated to be about 0.05 pixel.

### 2.4 Box counting method

Due to the discrete nature of the particle phase, a box counting method was used to obtain the spatial distribution of the statistics of the sand-grain motion. Following Kulick et al. (1994) and Aliseda et al. (2002), the whole FOV domain (0.80×0.16 m) was divided into a set of sub-boxes, each of which had size of $\Delta x \times \Delta y = 10 \times 1$ mm (or $\Delta x^+ \times \Delta y^+ = 175 \times 17.5$) with overlap ratio of 50%. The particle local volume fraction in each sub-box is calculated as

$$\Phi_V(x,y,t) = \frac{\pi d_p^3 N(x,y,t)}{6\Delta x \Delta y \Delta z}, \qquad (0.6)$$

where $N(x, y, t)$ is the number of identified sand grains within one sub-box centering around $(x, y)$ at time $t$, and $\Delta z$ is the laser-sheet thickness. Similar to Li et al. (2012), the fluctuating velocity of sand grains is defined with respect to the ensemble-averaged value within the local sub-box, i.e.

$$\vec{v}_p{'}(x_i, y_i, t) = \vec{v}_p(x_i, y_i, t) - \vec{V}_p(x, y), \qquad (0.7)$$

in which $x_i$ and $y_i$ are the absolute coordinates of the $i^{th}$ sand grain contained in the sub-box indexed by $(x, y)$, $\vec{v}_p$ is its instantaneous velocity and $\vec{V}_p$ the mean velocity averaged over all sand grains within this sub-box.

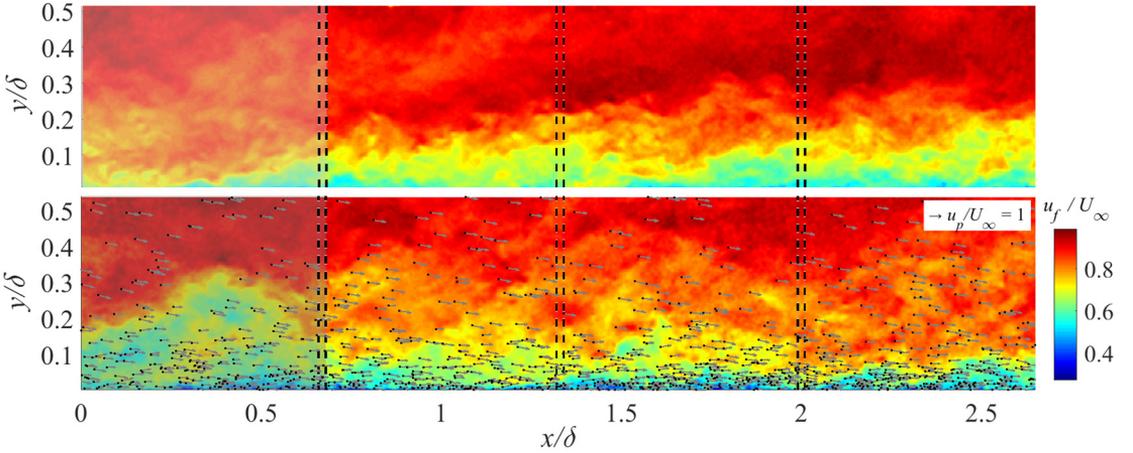

Figure 5. Instantaneous $u$-component velocity fields of the gas phase in $x$-$y$ plane. (a) Case S; (b) Case T1. The shaded area shows the FOV of an individual CCD camera. The black dots and gray arrows in (b) indicate sand grains and their translation velocities.

## 3. Partition of the sand motion

Figure 5 shows typical snapshots of instantaneous streamwise velocity fields $u_f(x, y)$ of the gas phase in both Case S and Case T1, in the latter of which the detected sand grains and their velocity vectors are also presented. Figure 5 leads to several interesting observations. First of all, the present multi-camera configuration is enough to accommodate the streamwise extend of LSMs. This can be quantified by the pre-multiplied spectrum $k_x\Phi_{uu}(k_x)$ of the streamwise velocity component shown in figure 6a. Although the limit streamwise extend of the FOV, i.e. $\Delta x=2.67\delta$, casts a truncation effect by compressing the large-scale side of the $k_x\Phi_{uu}(k_x)$ spectrum, one can still get an impression on the energy contribution of LSMs by the existence of a spectrum patch around $\lambda_x/\delta \approx 1$ across the whole wall-normal span.

Secondly, it can be inferred from figure 5 that the presence of sand grains does not change the strength and coherence of LSMs significantly. This can be evidenced by a comparison of large-scale part of $k_x\Phi_{uu}$ between Case S and Case T1/T2 in figure 6a. Nevertheless, the length scales of small-scale coherent motions in the near-wall region, which constitute the inner-layer peak in both $k_x\Phi_{uu}$ and $k_x\Phi_{vv}(k_x)$, get distinctly reduced. This aspect will be discussed in detail in §4.

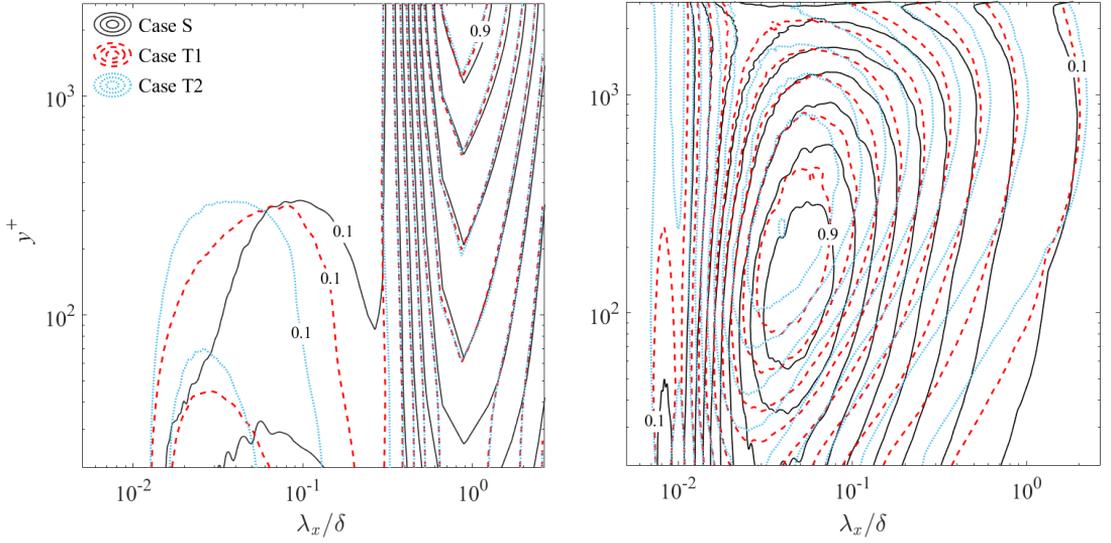

Figure 6. Pre-multiplied spectra of the fluctuation velocity of the gas phase in three tested cases. (a) streamwise velocity spectrum $k_x\Phi_{uu}$; (b) wall-normal velocity spectrum $k_x\Phi_{vv}$.

Thirdly, sand grains are non-uniformly distributed and particle concentration in the near-wall region is comparably higher than that in the outer layer. The following discussion on the time-averaged particle local volume fraction will give a quantitative support for this observation. Lastly, due to the gravity effect, most of sand grains present a downward motion with vertical velocity $v_p<0$, but ascending particles with $v_p>0$ are still presented, especially in the near-wall region. It indicates a remarkable saltation process, which might be related with the interaction between sand grains and near-wall turbulent cycles (Marchioli and Soldati 2002; Hout 2011, 2013), and will be further discussed in §5.

Recalling that sand grains, which were released from the wind tunnel ceiling at the test section inlet, descended towards the wall before entering into the measurement domain 8 m in the downstream. This means that the gravity plays a non-negligible role in affecting the kinematics and dynamics of sand grains, if comparing to the pure suspension scenario where the fluid-to-particle density ratio was close to unity (Hout 2011, 2013). In fact, as shown in figure 7(a), the mean streamwise velocity of sand grains in Case T1 and T2 is always larger than that of the gas phase at the same position, i.e. $U_p>U_f$, except for the region very close to the wall ($y/\delta=0.01$), where a remarkable streamwise deceleration makes $U_p$ lower than $U_f$ at the end of the FOV. Such a non-parallel behavior in the near-wall region is also presented in the streamwise variation of local particle concentration. As shown in figure 7(b), time-averaged particle local volume fraction $\Phi_V$ at $y/\delta=0.01$ monotonically increase with $x$, in distinct contrast to the quasi-constant $\Phi_V$ in higher flow layer.

To quantify the level of the streamwise non-homogeneity of the sand-grain kinematics, the streamwise gradients of $U_p$ and $\Phi_V$ over a large interval of $\Delta x=2\delta$ are calculated for each flow layer. As shown in figure 8, for both two particle-laden cases, the quick drop of the magnitudes of $\Delta\Phi_V/\Delta x$ and $-\Delta U_p/\Delta x$ with respect to $y$ ceases sharply at $y/\delta=0.12$, above which the magnitudes asymptotically converge to zero. This indicates a two-layer partition of the sand-grain behavior that is seldom observed in previous studies. Figure 9 gives a comparison of the histogram of the nominal size of sand-grain images in these two partitions in Case T1. Note that the nominal diameter was identified in the PTV algorithm, which served as an indirect measure of the physical size of sand grains. It is clearly shown that large (and heavy) sand grains have higher probability to reside in the near-wall region (lower than $y/\delta$=0.12), while those with smaller size (and lighter weight) tend to locate in the outer layer (higher than $y/\delta$=0.12).

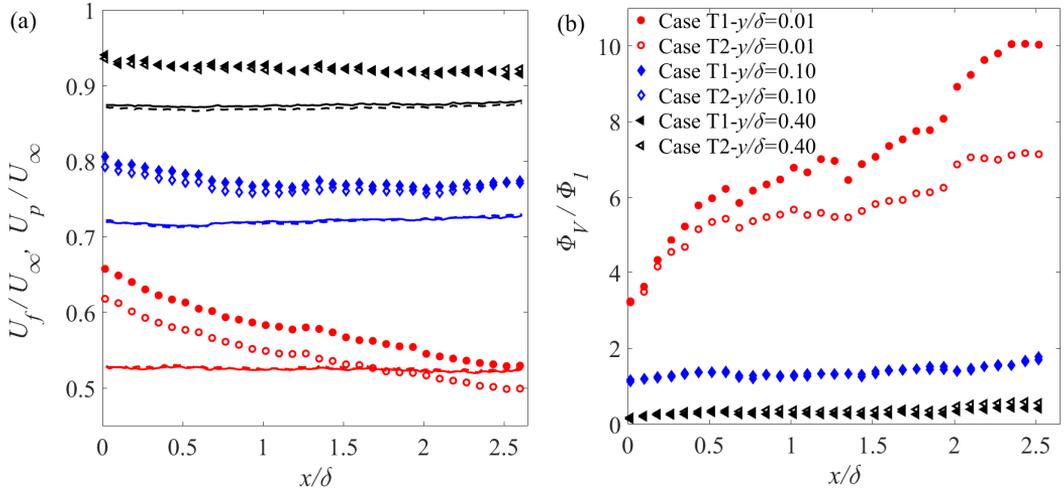

Figure 7. Streamwise variation of (a) the time-averaged streamwise velocity $U_f$ and $U_p$ of the gas and particle phase, and (b) the time-averaged particle local volume fraction $\Phi_V$ at three wall-normal heights in Case T1 and T2. In (a), three pairs of straight (Case T1) and dotted (Case T2) lines denote $U_p$ of the gas phase at $y/\delta$=0.01 (bottom), 0.10 (middle) and 0.40 (above), respectively. In (b), $\Phi_V$ is normalized by the particle bulk volume fraction $\Phi_1$ given in Table 3.

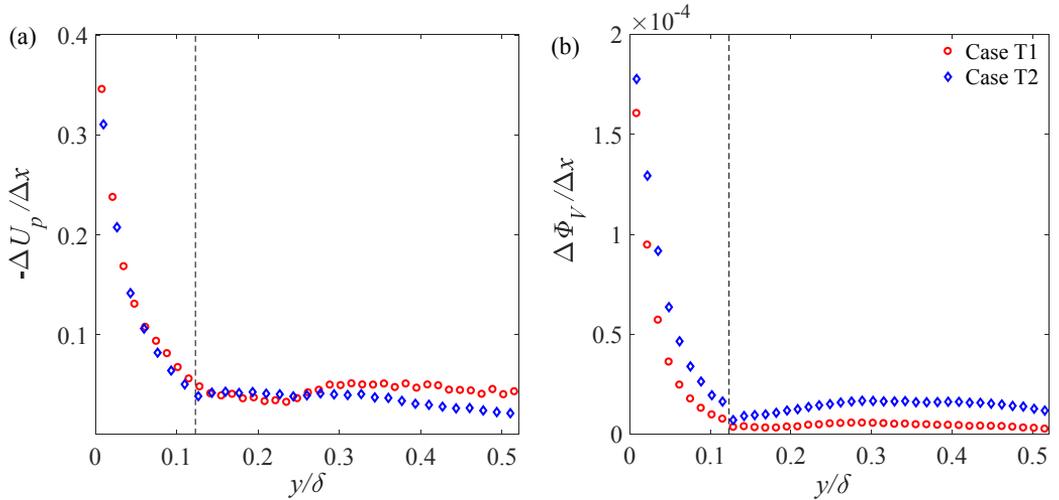

Figure 8. Wall-normal variation of (a) the streamwise gradient of the time-averaged particle local volume fraction $\Delta\Phi_V/\Delta x$, and (b) the streamwise gradient of the sand-grain mean velocity $-\Delta U_p/\Delta x$ in Case T1 and T2. The dotted vertical lines indicate the critical layer of $y/\delta$=0.12 (or $y^+$=670).

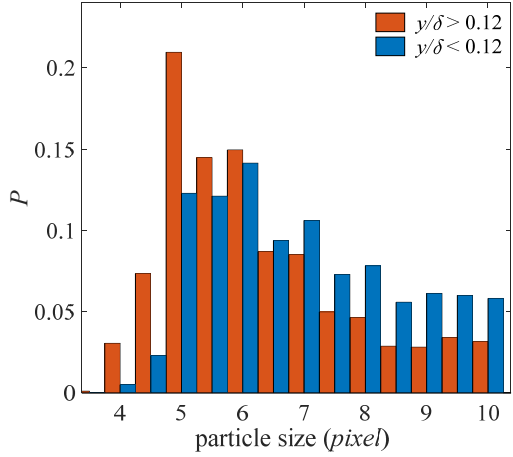

Figure 9 Distribution of the nominal diameter of sand-grain images in the above or below the critical layer of $y/\delta=0.12$ in Case T1.

According to the above observation, it can be inferred that the sand-grain motion in the outer layer is in a pseudo-equilibrium state. Specifically, light sand grains majorly present wall-parallel translation with minor vertical sedimentation, while the descending heavy sand grains towards the wall are compensated by the quasi-constant sand-grain flux from higher layer. On the other hand, the spatial developing nature of the sand-grain motion in the near-wall region might be associated with the particle saltation process due to the combined effect of inter-particle collision, particle-wall interaction and induction of inner-layer turbulent cycle (Zheng 2009; Wang and Zheng 2015). This leads to a continuous reduction of the sand-grain kinetic energy (see figure 7a) and a persistent enhancement of the particle accumulation (see figure 7b) in the downstream. Nevertheless, the reason why the critical layer separating these two partitions is $y/\delta=0.12$ (or $y^+=670$ in inner-scaling) is still unclear here, and will be further discussed in §4.

## 4. Two-phase velocity statistics

This section discusses both the turbulence modulation and the sand-grain motion in a sense that the gravity effect is non-negligible. Owing to the quasi-parallel behavior of sand grains in the outer layer, only the velocity statistics of the gas- and particle-phase at the end of the FOV, i.e. $x/\delta=2.5$, will be presented in the following. It should be noted that in the near-wall region the sand-grain statistics at $x/\delta=2.5$ cannot be interpreted as a representative of the whole FOV span.

### 4.1 *Turbulence modification*

Figure 10 summarizes the wall-normal profiles of the first- and second-order velocity statistics of the gas and particle phase in three tested cases. Comparing to the baseline Case S, the mean streamwise velocities of the gas phase $U_f$ in Case T1 and Case T2 get increased in the buffer layer ($y^+<60$) and are slightly attenuated in the wake region beyond $y^+\approx800$, while the log-law region remains nearly unchanged (see figure 10a). The velocity fluctuation intensities $u_{rms}$ and $v_{rms}$, on the other hand, are significantly attenuated in the inner layer with $y^+<670$ (see figure 10c, d), and the attenuation level seems to be positively correlated with the particle bulk volume fraction $\Phi_1$. Beyond the log-layer $u_{rms}(y)$ plateau, which is an indicator of the appearance of the so-called high-$Re$ effect in wall-bounded turbulence (Smits et al. 2011), the profiles of

$u_{f,rms}(y)$ of Case T1 and T2 gradually converge to those of Case S, representing an insignificant sand-grain effect due to the relative small particle concentration there.

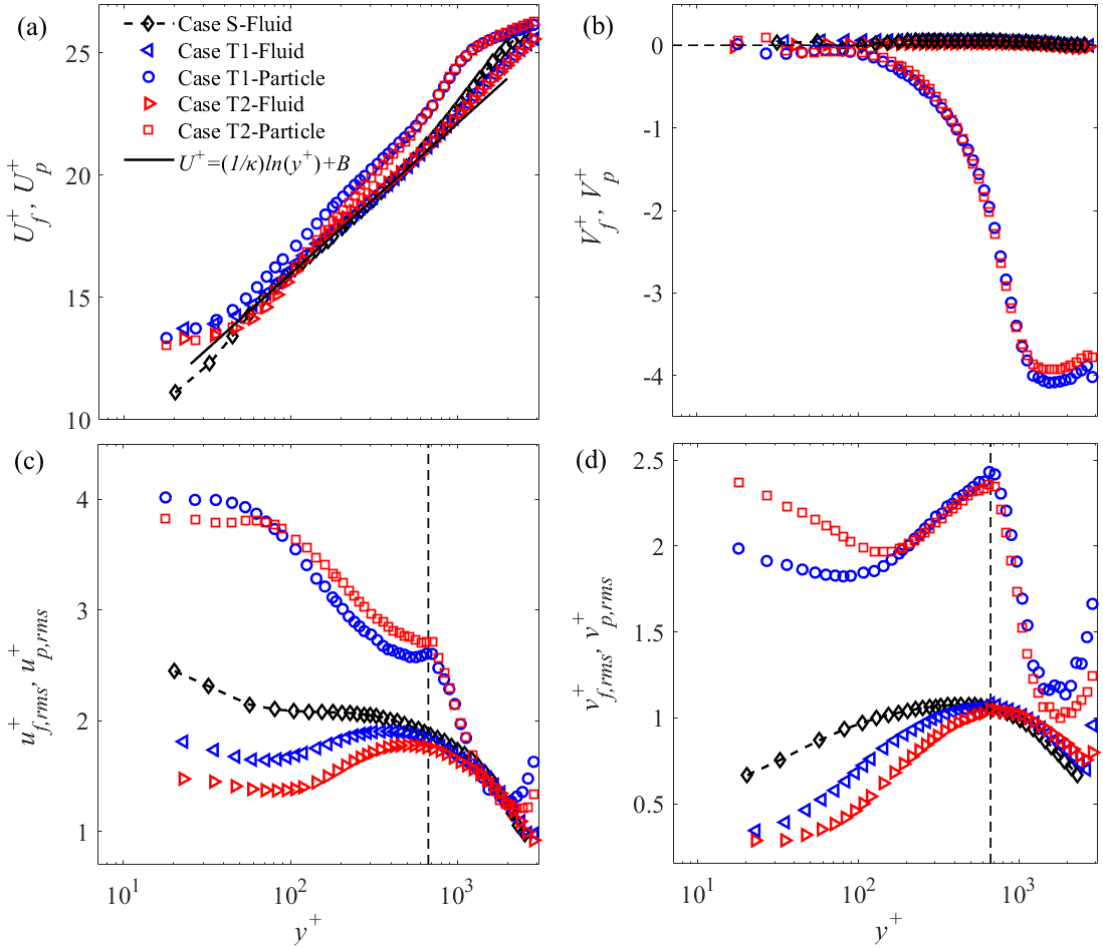

Figure 10 Wall-normal variation of the inner-scaled first- and second-order velocity statistics of the gas and particle phase at $x/\delta=2.5$ in three tested cases. (a) streamwise mean velocity $U^+$; --- $U^+=(1/\kappa)\ln(y^+)+B$ with $\kappa=0.384$ and $B=4.12$; (b) wall-normal mean velocity $V^+$; (c) streamwise velocity fluctuation intensity $u_{rms}^+$; (d) wall-normal velocity fluctuation intensity $v_{rms}^+$. The black dashed lines in (c) and (d) indicate the critical layer of $y/\delta=0.12$ (or $y^+=670$).

The observed turbulence attenuation in the near-wall region is consistent with recent experiments (Kiger and Pan 2002; Li et al. 2012) and simulations (Li et al. 2016, 2018). To our regards, two major factors among all the possible ones are responsible for the mechanisms of such turbulence attenuation. Firstly, sand grains with size much larger than the Kolmogorov length scale $\eta$ cause extra turbulence dissipation due to the presence of particle drag (Balachandar and Eaton 2010; Saber et al. 2016). Secondly, the crossing trajectory effect, induced by heavy particles falling out from turbulent eddies (Csanady 1963), is strengthened by large slip velocity between two phases (see figure 7a). Li et al. (2012) suggested that the crossing trajectory effect might contribute to the reduction of the spatial scale of turbulent coherent structures. It will be shown in §5 that the presence of sand grains suppresses the sweep and ejection events in the near-wall region, providing a side support for this effect.

In addition to these two factors, the gravity effect, which directly affects the kinematics of the sand grains, also plays a role in affecting those of the gas phase. This can be seen from both the slightly increase of $v_{f,rms}$ in the outer layer and the remarkable elevation of $U_f$ in the near-wall region (see figure 10a and d), which suggests a non-negligible exchange of the turbulent kinematic energy and the momentum from sand grains to the gas phase.

### *4.2 Sand-grain kinematics*

As for the particle phase in Case T1 and Case T2, the profiles of the mean streamwise velocity (shown in figure 10a) are always higher than those of the gas phase, i.e. $U_p > U_f$. The higher flow layer the larger deviation. This leads to a slightly steepening of the slope of $U_p(y)$ in the log-law region, which becomes more prominent in Case T2 with larger $\Phi_1$. In contrast, Kiger and Pan (2002) and Luo et al. (2017) reported that in wall-bounded flows the mean velocity profile of the gas phase and the particle phase hold the same slope in the log layer. The descending motion of the sand grains, on the other hand, is clearly revealed by the negative mean vertical velocity $V_p$ (see figure 10b). The magnitude of $V_p$ monotonically decreases as the sand grains descend towards the wall. Such wall-normal variation of $U_p$ and $V_p$ indicates a strong interaction between the particle phase and the gas phase, which gradually deflects the velocity vectors of the sand grains along the streamwise direction and counteracts the gravity effect to dissipate their mean kinetic energy.

Figure 10(c, d) show that the velocity fluctuation intensities of the particle phase in Case T1 and T2 are always larger than those of the gas phase, i.e. $u^+_{p,rms} > u^+_{f,rms}$ and $v^+_{p,rms} > v^+_{f,rms}$. This is consistent with previous studies on particle-laden channel flows (Kaftori et al. 1995; Tanière et al. 1997; Li et al. 2012; Coletti et al. 2016), and can be attributed to the combined effect of the gravity and the turbulent flow. Specifically, the gravity provides an additional source for the overall level of the TKE of the sand grains, while the turbulent flow spreads the enhanced TKE in a wide spectrum. To our regards, such a scenario remarkably shapes the profiles of $u^+_{p,rms}(y^+)$ and $v^+_{p,rms}(y^+)$. As shown in figure 10(c, d), a local minimum appears in both profiles in the wake region at about $y^+=2000$ ($y/\delta \approx 0.36$). In the below, both $u^+_{p,rms}$ and $v^+_{p,rms}$ quickly increase with the decrease of $y$, indicating the onset of the joint gravity-turbulence effect. This quick increase stops at the critical layer ($y^+=670$ or $y/\delta=0.12$) discussed in §3, below which either a slower increase of $u^+_{p,rms}$ or a sharp decrease of $v^+_{p,rms}$ appear in the log layer from $y^+=670$ to $y^+=100$. Finally, $u^+_{p,rms}$ asymptotes to constant in the buffer layer and below, while a re-growth of $v^+_{p,rms}$ is presented there.

The observation that the profiles of $u^+_{p,rms}(y^+)$ and $v^+_{p,rms}(y^+)$ both present an inflection point at the critical layer, which partitions the wall-parallel condition of the sand-grain motion, is interesting. In the previous studies, only Li et al. (2012) reported a similar observation, but no further discussion was addressed. To our understanding, this issue can be explained by the following three manifolds.

Firstly, as shown in figure 10(c, d), the inflection point is roughly in accordance with the peak position of the $v^+_{p,rms}$ profile of the gas phase. In addition, the center of the log-layer bump in the $u^+_{f,rms}$ profile, which originates from the log-layer $u_{rms}$ plateau due to the near-wall turbulence attenuation, is next to this position on the left side. Such a spatial accordance indicates a direct correlation between the sand-grain fluctuating motions and the large-scale turbulent structures in the log layer and below, which will be later evidenced in §5. Actually, figure 11(a) shows that a logarithmic decay of the local sand-grain volume fraction $\Phi_V$ with respect to $y$ is presented in the log layer of $y^+=60\sim670$. Note that the shape of the present profiles of $\Phi_V(y)$ is rather similar to those obtained by Pallares et al. (2014) and Coletti et al. (2016). Such a logarithmic distribution of the particle concentration was also observed by Creyssels et al. (2009) and Coletti et al. (2016). It implies that the attached eddies, which form the structural basis of the existence of the log layer in wall-bounded turbulence (Marusic et al. 2013; Baars et al. 2017), are responsible for the lateral convection and vertical motion of the sand grains in this region.

Secondly, the $u^+_{f,rms}$ profiles of Case T1 and T2 are seen to collapse well with that of Case S once beyond the critical layer (see figure 10c). This implies that due to the dilute nature of the local particle concentration (see figure 11a), the feedback effect of the sand grains on the gas phase is rather weak in the outer layer.

Lastly, the re-growth stage of $v^+_{p,rms}(y^+)$ below $y^+=100$ is in distinct contrast to the continuous decay of $v^+_{f,rms}(y^+)$. It provides a strong support for the inference proposed in §3; namely, the factors of inter-particle collision, particle-wall interaction and induction of inner-layer turbulent cycle contribute to the non-parallel behavior of the sand-grain kinematics in the log region and below.

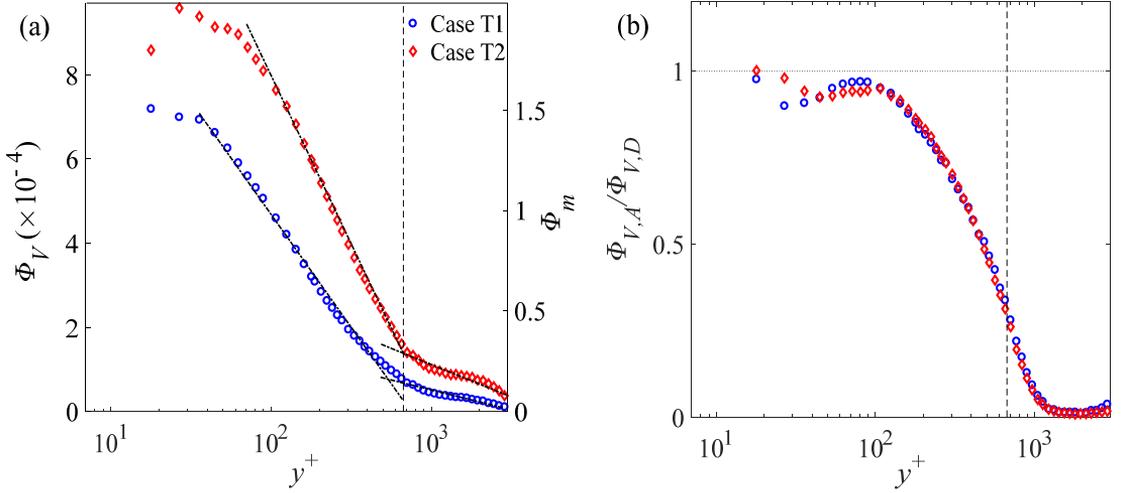

Figure 11 Wall-normal variation of (a) the particle local volume fraction $\Phi_V$ and (b) the ratio of local volume fraction of the ascending sand grains $\Phi_{V,A}$ to that of the descending ones $\Phi_{V,D}$ at $x/\delta=2.5$ in Case T1 and T2. The black dashed lines indicate the critical layer of $y^+ = 670$ (or $y/\delta = 0.12$).

### 4.3 *Statistics of ascending and descending sand grains*

Figure 10(b) shows that $V_p$ asymptotes to zero below $y^+<100$. This should not be interpreted as the absence of the vertical motion of the sand grains in the near-wall region, since the magnitude of $v^+_{p,rms}$ there is rather high (see figure 10d). In fact, individual sand grains in the near-wall region frequently present ascending (with $V_p>0$) or descending (with $V_p<0$) motions with strong amplitude (see figure 5b). Figure 11(b) shows the ratio of the local volume fraction of the ascending sand grains to that of the descending ones, i.e. $\Phi_{V,A}/\Phi_{V,D}$, in Case T1 and T2. The occurrence probabilities of the ascending and descending motions are seen to be close to each other below $y^+=100$. This leads to the mutual cancellation of their contribution to the averaged $V_p$, which can be further evidenced by the mean vertical velocity profiles of the ascending and descending sand grains, i.e. $V_{p,A}(y^+)$ and $V_{p,D}(y^+)$, shown in figure 12(b).

Figure 12 compares the velocity statistics between the ascending and descending sand grains in Case T1 and T2, those of the whole sand-gain ensemble are also supplemented as a reference. In the wake region with $y^+>1000$, a distinct peak appears in the wall-normal profiles of $u^+_{p,rms,A}$ and $v^+_{p,rms,A}$ of the ascending sand grains at around $y^+=2000$, which is absent in the corresponding profiles of the descending sand grains. Recalling that the concentration of the ascending sand grains in the wake region is one-order smaller than that of the descending ones (see figure 11b). Whether the low sampling fact leads to a non-converged peak is unclear; nevertheless, the contribution of the ascending sand grains to the whole velocity statistics in the outer layer is rather small.

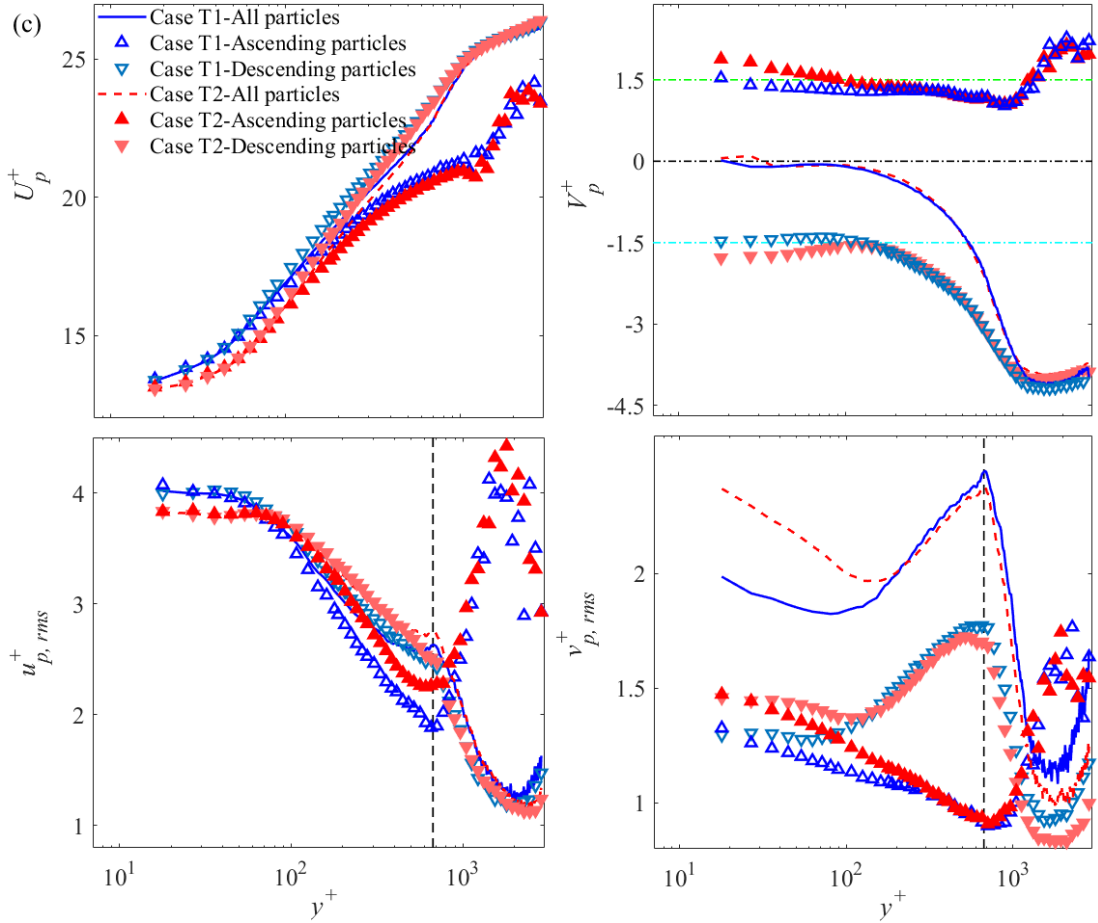

Figure 12 Wall-normal variation of the first- and second-order velocity statistics of the ascending and descending sand grains at $x/\delta=2.5$ in Case T1 and T2. (a) streamwise mean velocity $U_p^+$; (b) wall-normal mean velocity $V_p^+$; (c) streamwise velocity fluctuation intensity $u^+_{p,rms}$; (d) wall-normal velocity fluctuation intensity $v^+_{p,rms}$. The black dashed lines in (c) and (d) indicate the critical layer of $y^+=670$ (or $y/\delta=0.12$). The statistics of the whole sand grains are duplicated from figure 10 for comparison.

In the log layer, $U_{p,A}$, $u^+_{p,rms,A}$ and $v^+_{p,rms,A}$ of the ascending sand grains are always smaller than those of the descending ones. The observation of $U_{p,A}<U_{p,D}$ is consistent with the studies of Tanière et al. (1997) and Li et al. (2012). It indicates that the ascending sand grains gradually loss their kinetic energy in the saltation process. Nevertheless, such difference in $U_p$, $u^+_{p,rms}$ and $v^+_{p,rms}$ gradually diminishes below $y^+=100$.

Referring to the equivalent local volume fraction of the ascending and descending sand grains (see figure 11b), this highlights the role of the inter-particle collision that enhances the momentum transfer within sand grains in the near-wall region.

## 5. Correlation between turbulent structures and particle motions

The characteristics of the velocity statistics discussed in §4 implied strong correlation between turbulent coherent structures and sand-grain kinematics. This section will use both quadrant analysis and conditional-averaged method to provide direct evidence for such correlation, with focuses on the role of both the bursting events and LSMs in affecting the sand-grain vertical transport.

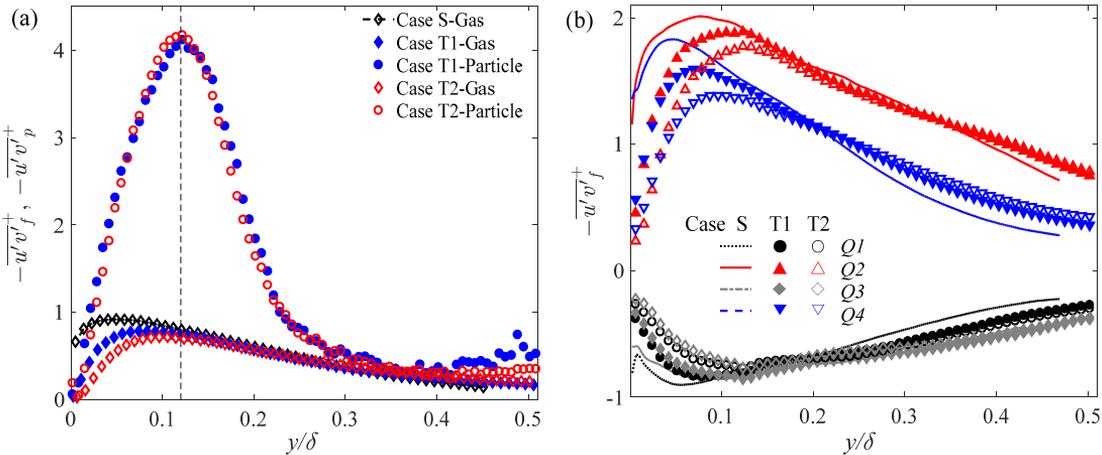

Figure 13 Wall-normal variation of (a) RSS of the gas and particle phase and (b) quadrant decomposed RSS of the gas phase at $x/\delta = 2.5$ in three tested cases. The black dashed line indicates the critical layer of $y/\delta = 0.12$ (or $y^+ = 670$).

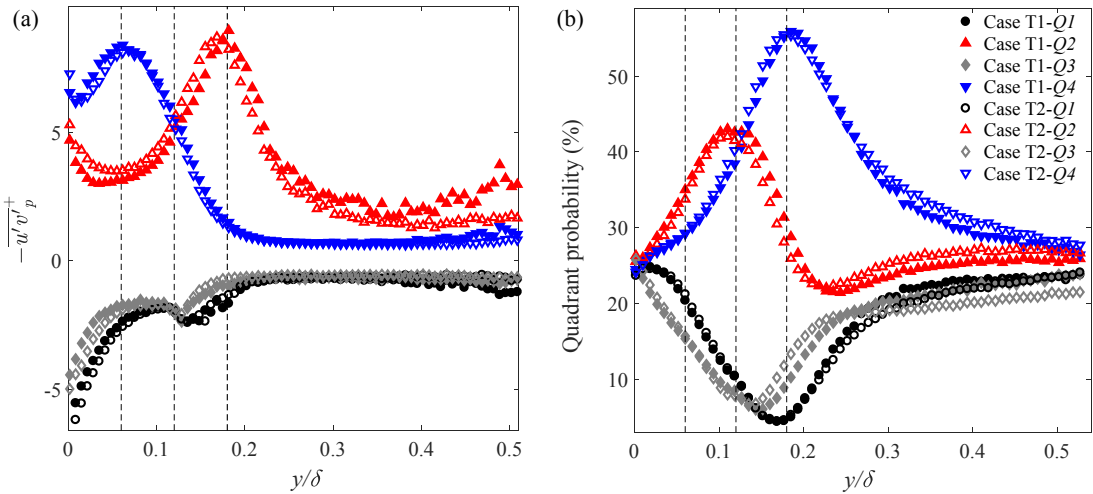

Figure 14 Wall-normal variation of (a) quadrant decomposed RSS contribution and (b) the occurrence probability of quadrant events of the particle phase at $x/\delta=2.5$ in Case T1 and T2. Three black dashed lines indicate $y/\delta= 0.06$, 0.12 and 0.18, respectively.

## 5.1 Quadrant analyses of the kinematics of the gas and particle phase

Figure 13(a) shows that comparing to Case S, the addition of the sand grains in Case T1 and T2 remarkably reduce the magnitudes of RSS of the gas phase, i.e., $-\overline{u'v'_f}$, in the log layer and below, the larger particle bulk volume fraction the higher RSS reduction level. This is consistent with the turbulence attenuation discussed in §4.1, and can be further explained by the quadrant-decomposed RSS profiles shown in figure 13(b), in which the reduction of the RSS is seen to be mainly attributed to the dampening of the Q2 and Q4 events of the gas phase in the inner layer.

On the other hand, the RSS of the sand grains significantly differs from those of the gas phase. As shown in figure 13(a), the $-\overline{u'v'_p}$ profiles peak at the critical layer of $y/\delta=0.12$ (or $y^+=670$), far beyond the peak position ($y/\delta=0.05 \sim 0.08$) of the $-\overline{u'v'_f}$ profiles, which slightly shifts towards higher layer with the increase of the particle bulk volume fraction. Moreover, the maximum values of $-\overline{u'v'_p}$ are three times larger than those of the gas phase. On considering that the inflection points in the $u_{p,rms}^+$ and $v_{p,rms}^+$ profiles are in accordance with the critical layer (as discussed in §4.2), the collapse of the peak position of the particle-phase RSS with the critical layer is not surprising. However, figure 14(a) shows that this RSS peak is actually a joint contribution of the Q2 and Q4 events of the sand-grain motions, whose individual contribution to the RSS reach maximum at $y/\delta \approx 0.18$ and 0.06, respectively. In contrast, the RSS peaks of the Q2 and Q4 events of the gas phase locate at $y/\delta=0.07\sim0.12$ and $y/\delta=0.04\sim0.08$ (see figure 13b), respectively. Figure 14(b) further shows that the local occurrence probability of the sand-grain Q2 events peaks at about $y/\delta \approx 0.12$, where the sand-grain Q2 and Q4 events have comparable occurrence probabilities, and their contributions to the overall RSS are almost the same.

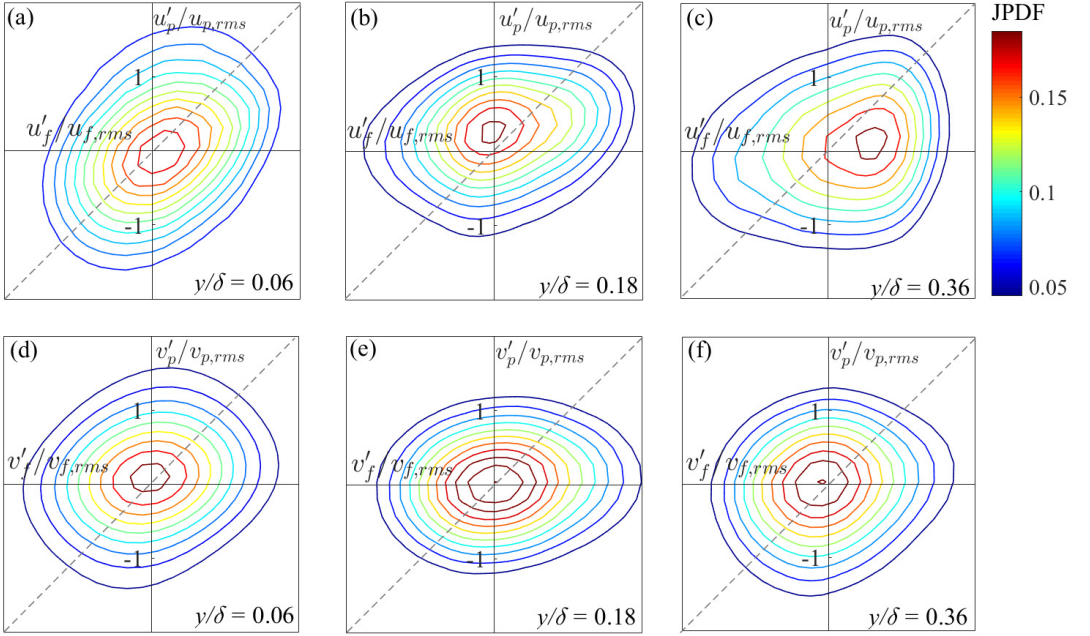

Figure 15 JPDF of $u'_f$ and $u'_p$ in (a-c) and $v'_f$ and $v'_p$ in (d-f) at $x/\delta=2.5$ and different wall-normal height in Case T1. (a, d) $y/\delta=0.06$; (b, e) $y/\delta=0.18$; (c, f) $y/\delta=0.36$. The instantaneous fluctuation values are normalized by their corresponding local fluctuation intensity. The dashed diagonal line in each panel indicates a full correlation between two studied variables.

Such difference indicates that the sand grains are not merely a passive tracer to follow the turbulent burst events. This can be further evidenced by the joint probability density functions (JPDFs) of the fluctuating velocity $u'_f$ (or $v'_f$) and $u'_p$ (or $v'_p$) of the gas and particle phase at the same time and the same space. Similar to Kiger and Pan (2002), Hout (2011) and Li et al. (2012), $u'_f$ and $v'_f$ at the position of one sand grain is actually interpolated from 16 neighboring gas-phase velocity vectors surrounding this sand grain via bi-cubic interpolation. In Case T1, the JPDF at $y/\delta=0.06$ (shown in figure 15a and d) presents an elliptical shape with the major axis inclined at 45° with respect to the abscissa axis. It suggests a strong correlation between the sand-grain kinematics and the turbulent motions in the near-wall region, which will be further explored in §5.2. Nevertheless, this correlation gets relaxed in the wake region, where the slope of the major axis of the JPDF gradually reduces (see figure 15b, c, e and f with $y/\delta=0.18$ and 0.36, respectively). Note that such trend is also observed in Case T2.

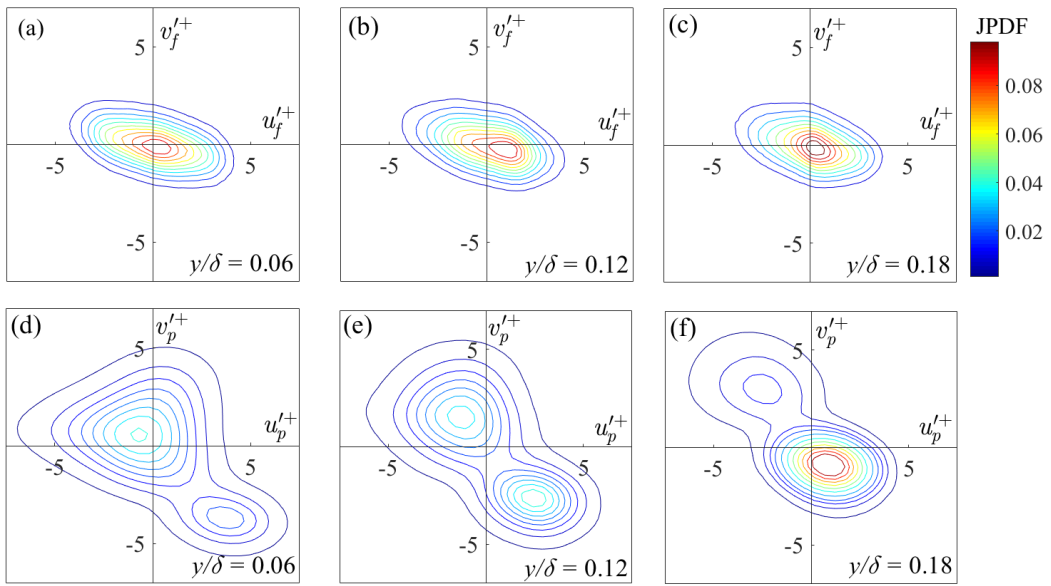

Figure 16 JPDF between $u'_f$ and $v'_f$ of the gas phase in (a-c) and $u'_p$ and $v'_p$ of the sand grains in (d-f) at $x/\delta = 2.5$ and different wall-normal height in Case T1. (a, d) $y/\delta=0.06$; (b, e) $y/\delta=0.12$; (c, f) $y/\delta=0.18$.

Figure 16 further shows the JPDFs of the 2D fluctuating velocity components of both the gas phase and the particle phase, i.e. $JPDF(u'^+_f, v'^+_f)$ and $JPDF(u'^+_p, v'^+_p)$, at $y/\delta=0.06$, 0.12 and 0.18 in Case T1. Note that Case T2 presents similar JPDFs and is not presented for concise. Unlike the elliptical-shaped $JPDF(u'^+_f, v'^+_f)$ of the gas phase, $JPDF(u'^+_p, v'^+_p)$ of the particle phase always presents two distinct peaks locating in the second and fourth quadrants, respectively. This is a strong evidence for the co-existence of the ascending and descending motions of the sand grains. However, the weakening trend of the Q2 events of the particle phase in higher flow layer is also presented, consistent with the observation in figure 14(b) that the occurrence frequency of the descending Q4 events becomes dominant beyond the critical layer of $y/\delta=0.12$. Therefore, it can be now inferred that the critical layer actually reflects a balance between the Q2 and Q4 events of the sand grains. Below the critical layer, the particle saltation process is mainly associated with the near-wall turbulent ejection events, thus contributing to the spatial developing nature of the sand-grain kinematics there; while in the above, it is the gravity effect, instead of the turbulent sweep events, that dominates the particle descending motions.

### 5.2 *Effect of large-scale turbulent structures on particle motions*

To give a direct description on the correlation between the descending and ascending motions of the sand grains and the turbulent flows, the technique of conditional average, which have been widely used in the study of turbulent coherent structures (Ganapathisubramani et al. 2012; Pallares et al. 2014; de Silva et al. 2018; Jiménez 2018), is conducted to reveal the averaged flow structures that are associated with the sand-grain Q2 and Q4 events with strong amplitude. With this purpose, the probing condition is set as

$$u'_p < -\sigma_{u'_p}, v'_p > \sigma_{v'_p} \qquad (5.1)$$

for Q2 events and

$$u'_p > \sigma_{u'_p}, v'_p < -\sigma_{v'_p} \qquad (5.2)$$

for Q4 events, respectively, where $\sigma_{u'_p}$ and $\sigma_{v'_p}$ are the sand-grain velocity fluctuation intensity at the probed flow layer $Y_p$. Such a condition excludes weak quadrant events that contribute less to the RSS of the particle phase. The instantaneous flow fields of the gas phase centering around the detected sand grains in Q2 or Q4 events are extracted, aligned and then averaged.

Figure 17(a, b) illustrates the conditional-averaged gas-phase velocity fields associated with the extreme Q2 and Q4 events of the sand grains being probed at $Y_p/\delta=0.12$ in Case T1. The conditional-averaged flow field at other $Y_p$ or in other case (Case T2) have similar pattern and are not shown for concise. The strong spatial correlation of the Q2 and Q4 events between the sand grains and the turbulent flow is clearly seen, consistent with previous studies (Kiger and Pan 2002; Marchioli and Soldati 2002; Vinkovic et al. 2011; Hout 2013). More interestingly, owing to the large FOV of the present measurement, it is clearly shown (in figure 17) that the gas-phase Q2 or Q4 events surrounding the sand grains actually present a shape similar to low- or high-speed LSMs with streamwise length scale larger than $2\delta$. To our knowledge, this is, for the first time, a direct evidence for the concept that the sand-grain motions are more or less affected by large-scale turbulent structures (Wang et al. 2017).

Nevertheless, figure 17b reveals a non-negligible phase difference in the Q4 events between the descending sand gain and the high-speed LSM. That is, the spatial position of the probed sand grain in Q4 event (white dot in figure 17b) locates in the downstream and below of the center of the high-speed LSM, which is measured by the local peak in the conditional-averaged $u$ component field (black cross in figure 17b). In contrast, an acceptable phase accordance between the ascending sand grains and the centers of the low-speed LSMs is seen for the Q2 event (see figure 17a) at the same probing flow layer. Figure 18 further shows that such phase relationship holds in the whole wall-normal FOV span. Specifically, those sand gains in extreme Q4 motion prefer to lie below the high-speed LSMs, while those Q2-typed sand gains usually follow the convection of low-speed LMSs, except for the near-wall region (with $y/\delta<0.06$) where they present a remarkable phase lag in the streamwise direction.

To our regards, the observation that the Q4-typed descending sand grains seem to be pushed by high-speed LSMs is actually a result of the gravity effect, which elevates the descending velocity of the Q4-typed motion of the sand grains triggered by the turbulent sweep event. Meanwhile, the streamwise lag of the Q2-typed ascending sand grains in the near-wall region highlights the effect of the sand grains' inertia on the particle saltation process. It is noted that neither of these two phenomena was reported in the previous studies. They deserve to be further studied for a fully characterization of the kinematics of heavy particles in particle-laden turbulent flows.

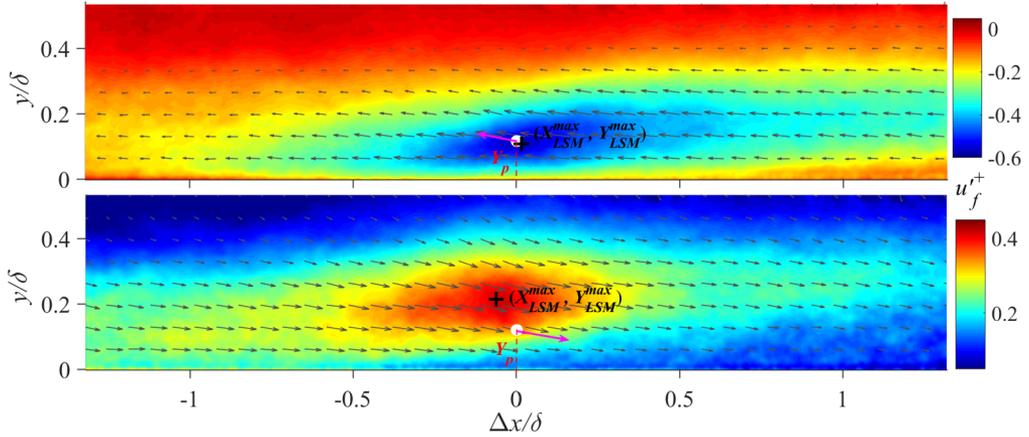

Figure 17 Conditional averaged velocity field of the gas phase subject to the condition of (a) extreme Q2 event and (b) extreme Q4 event of the sand grains being probed at the flow layer of $Y_p/\delta=0.12$ in Case T1. The white dots indicate the location of the probed sand grains, and the black crosses indicate the center of the LMSs in the gas phase whose coordinates are denoted as $(X_{LSM}^{max}, Y_{LSM}^{max})$. The gray arrows are the conditional averaged velocity vectors, the background contours are the distribution of $u'^{+}_f$, and the magenta arrows denote the Q2 or Q4 event of sand grains with no indication of the event strength.

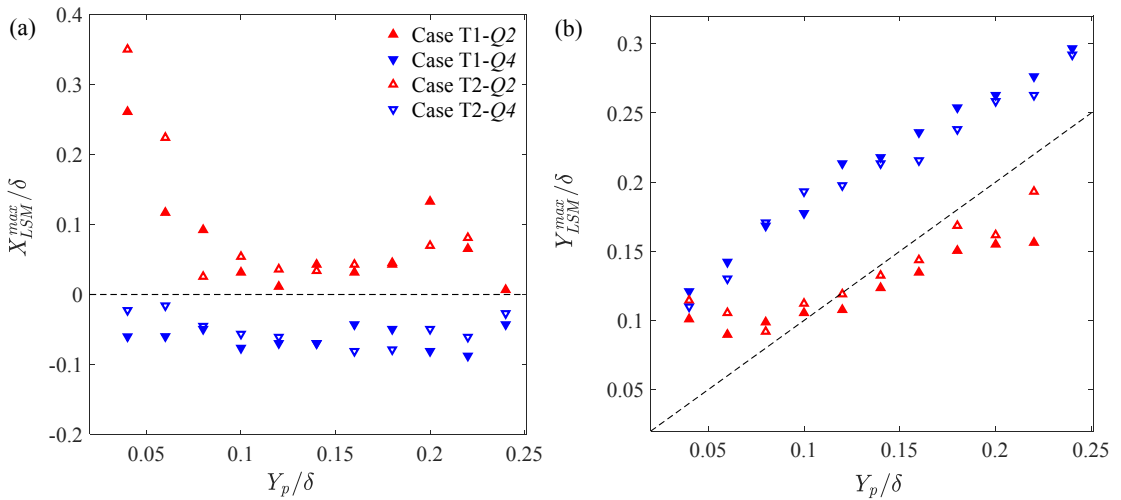

Figure 18 Spatial position of the center of LSMs surrounding either Q2- or Q4-typed sand grains as a function of the probing position $Y_p$. (a) $X_{LSM}^{max} \sim Y_p$; (b) $Y_{LSM}^{max}/\delta \sim Y_p$. The dashed lines indicates a full phase accordance between the sand grains and the LSMs.

## 6. Discussion and concluding remarks

Using joint PIV/PTV technique to obtain a large field-of-view two-dimensional measurement, turbulence modification and particle transport are investigated in a particle-laden TBL at relatively high $Re$. Heavy sand grains with remarkable inertia are used as the dispersed particle phase. After being released from the ceiling of the test section of a low-speed wind tunnel, they are translated by the free-stream flow over a considerably long distance before entering into the measured turbulent boundary layer. This configuration is used to simulate the particle deposition stage in the sand-storm flow, in which the effect of the gravity is non-negligible.

The most interesting finding of the present work, in our regards, is the existence of a critical layer that partitions the streamwise evolution of both the local volume fraction and the streamwise mean velocity of the sand grains into a spatial developing near-wall region and a quasi-parallel outer region. Across this critical layer, the particle velocity fluctuation intensities are found to present a severe change. Meanwhile, the logarithmic decay of the particle local volume fraction also ceases beyond it. Quadrant analysis reveals that a balance of the occurrence probabilities between the ascending and descending sand grains is reached at this critical layer, where their contribution to the local RSS is also equivalent with each other.

Such a critical layer might be a good indicator of the upper bound of the particle saltation process, which results in the considerably large fraction of the ascending sand grains in the below of it. On considering the relative high position of this critical layer ($y/\delta$=0.12 or $y^+$=670), it is reasonable to question that the factors of particle-wall interaction and inner-layer bursting events themselves are not enough to sustain the ascending motion of heavy particles in the log layer. This can be evidenced by the fact that even in the log layer, the RSS strength of the Q2 events of the particle phase is much larger than that of the gas phase (see figure 13b and figure 14a).

The conditional average technique, in together with the large-FOV measurement, credits us the ability of exploring the relationship between the extreme quadrant events in the particle phase and the large-scale turbulent structures in the gas phase. This analysis reveals a strong spatial correlation between the particle-phase extreme Q2 events and the gas-phase low-speed LSMs in the log layer and below. It forms a new evidence for the concept that low-speed LSMs do play a significant role in lifting-up heavy sand grains in the particle saltation process. Specifically, due to both the Stokes drag and the gravity, sand grains in the near-wall region are usually dragged by low-speed LSMs in the behind (see figure 18). Such a scenario also explains the observation in figure 10(c) that the critical layer is next to the right end of the log-layer $u_{rms}$ plateau (or bump) in the TBL. According to Deng et al. (2018), the detachment of LSMs from the wall marks a quick drop of the magnitude of $u_{rms}$ and a cease of the $u_{rms}$ plateau. If the role of LSMs on the sand-grain vertical motion is valid, the detachment of LSMs will also lead to a significant weakening of the occurrence probability of the sand-grain Q2 events, which exactly occurs beyond the critical layer (see figure 14b). This will finally form the upper bound of the particle saltation process.

Turbulence attenuation, which is positively correlated with the particle bulk volume fraction, mainly confines in the near-wall region. This observation is similar to most of the previous studies on particle-laden flow with relatively large particle size. The new (and interesting) finding via spectrum analysis is that the small-scale part of the turbulent fluctuations in the TBL are more vulnerable to the attenuation effect (shown in figure 6), which embodies as the reduction of the strength of the inner-layer turbulent cycle in the buffer region and below (shown in figure 13b). Large-scale fluctuations, on the other hand, are less affected by the presence of the sand grains. On considering the scale ratio between large-scale turbulent structures and individual particles, this observation is reasonable. Such a difference in turbulence attenuation among different scales reminds us the necessary to take the length scale into consideration when differentiating the turbulence-particle coupling. Nevertheless, if the particle preferential concentration amounts to a significant level, can it affect the kinematics and dynamics of large-scale turbulent structures? This is an interesting issue to be studied in the future.

In summary, the present work provides a reasonable justification to show that large-scale turbulent structures in TBL is an indispensable factor in the particle saltation process that significantly contributes to the vertical flux of heavy particles. Whether this factor also contributes to the formation of sand storm and the transport of sand grains over long distance is still inconclusive. To fully answer this question, future work on the relationship between particle preferential concentration and large-scale turbulent structures in high and extreme high *Re* condition is needed.

## Acknowledgments

We gratefully acknowledge Jing Li in Huazhong University of Science and Technology for insightful discussions. Hang-Yu Zhu greatly appreciates the assistance of Di Wu, Guo-Hua Wang, Xiao-Cang Ji, Wei Zhu and Jie Zhang in the experiment. This work was financially supported by the National Natural Science Foundation of China (grant nos. 11490552, 11672020 and 11721202).